\documentclass[12pt]{iopart}

\usepackage{graphicx}
\usepackage{epsfig}

\usepackage{txfonts}

\begin{document}         

\title[Optical Tweezers in a Sagnac Interferometer]{Sagnac Interferometer Enhanced Particle Tracking in Optical Tweezers} 

\author{M.~A.~Taylor$^1$, J.~Knittel$^1$ , M.~T.~L.~Hsu$^{1, 3}$,  H.-A.~Bachor$^2$, and W.~P.~Bowen$^1$ \footnote[1]{To whom correspondence should be addressed wbowen@physics.uq.edu.au}}   
% \date{May 28, 2010}   

\address{$^1$\ School of Mathematics and Physics,
 University of Queensland, Brisbane 4072, Australia }

\address{$^2$\ Australian Centre for Quantum-Atom Optics, Department
of Physics, Australian National University, Canberra ACT 0200,
Australia}

\address{$^3$\ National Measurement Institute,
PO Box 264, Lindfield, NSW 2070, Australia}

\begin{abstract}

A setup is proposed to enhance tracking of very small particles, by using optical tweezers embedded within a Sagnac interferometer. The achievable signal-to-noise ratio is shown to be enhanced over that for a standard optical tweezers setup. The enhancement factor increases asymptotically as the interferometer visibility approaches $100\%$, but is capped at a maximum given by the ratio of the trapping field intensity to the detector saturation threshold. For an achievable visibility of $99\%$, the signal-to-noise ratio is enhanced by a factor of 200, and the minimum trackable particle size is 2.4 times smaller than without the interferometer.

\end{abstract}


\maketitle

\section{Introduction}

Optical tweezers are devices which trap and detect small particles in a tightly focused laser beam~\cite{Afzal}. Radiation pressure draws particles towards higher light intensities, and traps them in the focus of the beam. The effect of a particle on the trapping beam profile can be analyzed to extract information about the position of and force on the trapped particle with subnanometer and subpiconewton detection sensitivity~\cite{Lang,Neuman}. This has become an important technique in a range of applications, particularly high-precision manipulation of biological samples. Optical tweezers have been used to manipulate viruses and bacteria~\cite{Ashkin1987}, unfold single RNA molecules~\cite{Bustamante}, study the motion of the biological motor protein kinesin~\cite{Svoboda1993} and muscle myosin molecules~\cite{Finer}, and sequence DNA~\cite{Greenleaf}.

Particles trapped in optical tweezers are tracked via the interference between the trapping beam and the light which scatters off the particle~\cite{Gittes,Harada}. A quadrant photodiode can be used to infer both the position of the particle and the force exerted on it~\cite{Pralle,Simmons}. Detecting the particle becomes much more difficult as it becomes smaller, because small particles can scatter very little light, with the amplitude of Rayleigh scattering scaling as the particle radius to the power of six~\cite{Berne,Hulst}. Particles which have been successfully trapped and detected in optical tweezers include 26~nm dielectric particles~\cite{Ashkin1986} and 36~nm gold nanoparticles~\cite{Svoboda}. The sensitivity of such measurements is limited by low scattered light levels.

 In almost all cases, and especially when trapping small particles, the trapping field intensity used in optical tweezers is much brighter than the saturation threshold of the detector used to track particle position. In order to detect the particle, there are two options. A second beam which is much less bright may be added to the optical tweezers setup, and this beam can be used to detect the particle position. This second beam may be orthogonally polarized to the trapping beam~\cite{Sehgal}, or it may be at a different wavelength~\cite{Simmons}. Alternatively an attenuator is placed in the beam between the optical tweezers and the detector. This attenuates the trapping beam, but it also attenuates the light scattered from the particle, degrading our ability to detect the particle. The signal-to-noise ratios (SNRs) achieved with these two approaches are identical if we assume optical linearity throughout the system.

Several methods to improve the sensitivity of optical tweezers exist, using for example back focal plane interferometry~\cite{Allersma}, two orthogonally polarized beams~\cite{Svoboda1993}, or spatial homodyne detection~\cite{Hsu,Tay}. Sophisticated techniques have also been proposed to surpass the shot noise detection limit using quantum states of light~\cite{Treps2002,Treps2003,Treps2004}. Both spatial homodyne detection and the use of quantum states of light could be integrated into the technique proposed in this paper to further enhance the particle tracking capability.

This paper proposes the combined use of optical tweezers and Sagnac interferometry for enhanced particle tracking, extending a recent demonstration of Sagnac interferometer based phase plate characterization~\cite{TaySagnac}. With optical tweezers embedded in a Sagnac interferometer, selective interference attenuates the trapping field and hence reduces the detection shot noise, substantially improving the detection SNR when compared to using a standard attenuator.
The particle tracking SNR is enhanced by a factor which increases as the interferometer visibility approaches $100\%$, up to a maximum enhancement defined by the ratio of the trapping field power to the detector saturation threshold. 
If, for example, the Sagnac interferometer had a visibility of $99\%$, the signal-to-noise ratio would be  enhanced by a factor of 200, which would consequently enable tracking of 2.4 times smaller particles than the equivalent standard optical tweezers scheme.

\section{Theory}

\begin{figure}
 \begin{center}
   \includegraphics[width=6cm]{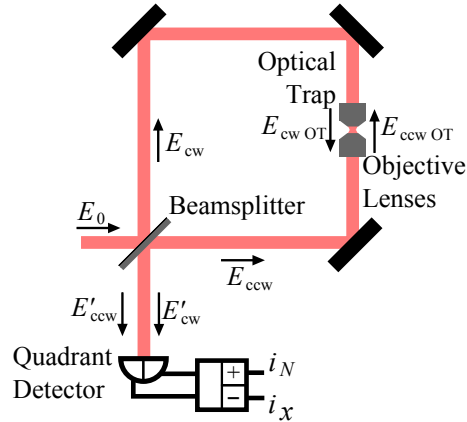}
   \caption{Layout of the Optical Tweezers detection scheme.  The trapping field is split at the beam splitter, with the transmitted field $E_{\rm ccw}$ traveling counterclockwise around the Sagnac interferometer, and the reflected field $E_{\rm cw}$ traveling clockwise. Once the fields reach the beam splitter again they recombine and interfere. The quadrant photodiode detects the light, producing sum and difference photocurrents $i_N$ and $i_x$. }
 \label{layout} 
 \end{center}

\end{figure}

A schematic of the  Sagnac interferometer embedded optical tweezers proposed here is shown in Fig.~\ref{layout}. The input optical field $E_0$ is split by a beam splitter, resulting in two optical fields propagating through the Sagnac interferometer, $E_{\rm ccw}$ traveling counterclockwise and $E_{\rm cw}$ traveling clockwise. These fields form an optical trap at the focus of the objective lenses. When a particle is trapped, it will scatter light from both fields, modifying their spatial profiles. The fields then recombine at the beam splitter, with the trapping field 
constructively interfering when returning out the beam splitter port of incidence, henceforth termed the {\it light port}, as is standard for a Sagnac interferometer. The quadrant detector used to extract particle position information is placed out the other {\it dark port}, where the trapping field destructively interferes. By contrast,
 the component of the scattered field containing particle position information constructively interferes when leaving the dark port, provided the interferometer has an odd number of internal reflections.

The number of interferometer mirrors is kept general in the following theory to illustrate the necessity for an odd number of internal reflections. Phase shifts upon hard boundary reflection from mirrors have no effect on the interference of the clockwise and counterclockwise fields, as both fields experience the same number of reflections. For simplicity we therefore neglect them. We have defined the {\bf \^{z}} axis in the direction of propagation of the laser beam, {\bf \^{y}} as normal to the plane of the interferometer, and {\bf \^{x}} by {\bf \^{x}}~=~{\bf \^{y}}~$\times$~{\bf \^{z}}, and we assume that the polarizations of all optical fields are the same throughout the experiment, so that all fields can be treated as scalar. We also assume that the input trapping field $E_0$ is symmetric on reflection, so $E_0(x,y) = E_0(-x,y)$.

 Since the clockwise and counterclockwise fields travel the same optical path in opposite directions through the setup, they will form a standing wave. At the trapping position the effect of this will be to greatly increase the light intensity gradient in the axial direction. Standing waves have previously been used in optical tweezers to improve the axial trapping of sub-wavelength particles~\cite{Zemanek}. Although not strictly necessary, for simplicity we assume that this is achieved, such that the particle is trapped in the {\bf \^{z}} direction at an antinode of the standing wave. This constraint on the axial position of the particle will ensure that light backscattered by the particle from the clockwise and counterclockwise trapping fields will have a phase such that it constructively interferes at the light port of the interferometer, and destructively interferes at the dark port. This allows backscattered light from the particle to be neglected in the following derivation.

Throughout this work it is useful to separate the electric fields into normalized real modes $u(x,y)$ and complex amplitudes $A$, such that
\begin{equation}
E_n(x,y) = A_n u_n(x,y),\\
\end{equation}
where $n$ is an arbitrary subscript, and $u_n$ is normalized such that
\begin{equation}
\iint \limits_{-\infty -\infty} ^{ \infty ~ \infty}  \! \! u_n(x,y)^2 dx \, dy =1. \label{mode}
\end{equation}

The transmittance and reflectance of the beam splitter are given by $T$ and $R$ respectively, so that the transmitted field $E_{\rm ccw}$ is given by
\begin{equation}
 E_{\rm ccw}(x,y) = \sqrt{T}E_0(x,y).
\end{equation}
This field propagates a distance of $L_1$ to the optical tweezers, and picks up a phase shift of $e^{i k L_1}$, where $k=\frac{2 \pi}{\lambda}$ is the wavenumber and $\lambda$ is the optical wavelength.
To first order, sufficiently small trapped particles leave the trapping field unchanged except for the introduction of a component $E_p$ which is scattered from the particle~\cite{Gittes}.
% To first order, and for sufficiently small particles, when incident on the trapped particle the optical tweezers leave the field unchanged except for the introduction of a component $E_p$ which is scattered from the particle \cite{Gittes}. 
After interaction with the particle, the field is
\begin{equation}
E_{\rm ccw~OT}(x,y) = \sqrt{T} E_0(x,y) e^{i k L_1}+  E_p(x,y) \label{E1eqn1}.
\end{equation}

 The scattered field $E_p$ can be separated without loss of generality into symmetric and antisymmetric parts $E_s$ and $ E_a$, so that
\begin{equation}
E_p(x,y) = E_s(x,y)+E_a(x,y),
\end{equation}
which due to their symmetry have the  properties
\begin{eqnarray}
 E_s(x,y) =  E_s(-x,y)\\
 E_a(x,y) = -E_a(-x,y) \label{asymmetry}.
\end{eqnarray}
 This is useful because all of the particle position information is found in the antisymmetric part of the scattered field for small particle displacements~\cite{Tay}. The amplitude of the scattered field is proportional to the counterclockwise trapping field amplitude, such that $A_s = \sqrt{T} A_0  e^{i k L_1}  \zeta_s$, and $A_a = \sqrt{T} A_0  e^{i k L_1} \zeta_a$, where $\zeta_s$ and $\zeta_a$ respectively denote the proportion of the trapping field which scatters from the particle into symmetric $u_s(x,y)$ and antisymmetric $u_a(x,y)$ modes. Here we work in the experimentally relevant limit that the proportion of the trapping field which is scattered is very small, or equivalently, $\{ \zeta_a, \zeta_s \} \ll 1$. Substituting these expressions into Eq.~(\ref{E1eqn1}), we find
\begin{equation}
E_{\rm ccw~OT}(x,y) = \sqrt{T} A_0 e^{i k L_1} [ u_0(x,y)+ \zeta_s u_s(x,y) + \zeta_a u_a(x,y) ] \label{E1eqn2}.
\end{equation}

\begin{figure}
 \begin{center}
   \includegraphics[width=5cm]{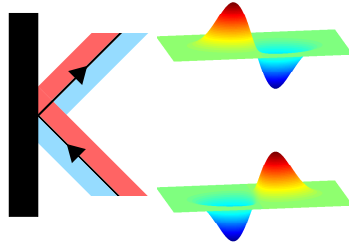}
   \caption{Phase induced by reflection of an antisymmetric field off a mirror. Left: reflection of an antisymmetric field off a mirror. Right: spatial profiles of field before and after reflection. The example spatial profile shown here is a TEM01 mode.}
 \label{reflection} 
 \end{center}

\end{figure}

Each reflection off a mirror causes a reflection of the beam profile in the $x$ direction. This is shown graphically in Fig.~\ref{reflection}. As seen in Eq.~(\ref{asymmetry}), this results in a change in the sign of the antisymmetric scattered field $E_a$, but does not effect either the trapping field $E_0$, which we assume to be symmetric as is typical in optical tweezers, or the symmetric scattered field $E_s$. As a result the antisymmetric scattered field, which contains the particle position information, picks up an additional phase shift on each reflection compared to the trapping field and symmetric scattered field.

Similar analysis of the clockwise path through the interferometer yields the clockwise optical field after interaction with the particle,
\begin{eqnarray}
E_{\rm cw~OT}(x,y) &=&  -\sqrt{R} E_0(x,y) e^{i k L_2}+ E_p(x,y)  \\
 &=&  -\sqrt{R} A_0 e^{i k L_2} [ u_0(x,y)+ \zeta_s u_s(x,y) + \zeta_a u_a(x,y) ]  ,
\end{eqnarray}
where $L_2$ is the distance traveled to the optical trap, and the negative sign is due to a hard boundary reflection from the beam splitter.

For the sake of brevity the explicit spatial dependence is now dropped, with $u_n(x,y)$ written as $u_n$ throughout. After interaction with the particle, both counterclockwise and clockwise fields propagate back to the beam splitter. The phase shift due to beam propagation cancels out because both beams travel the same total distance. The counterclockwise and clockwise fields experience $g$ and $f$ reflections respectively before reaching the beam splitter, with each reflection inducing a $\pi$ phase shift on their antisymmetric components. The fields at the beam splitter are then
\begin{eqnarray}
E'_{\rm ccw} = \sqrt{T} A_0 [\left( u_0+ \zeta_s u_s \right) + (-1)^g \zeta_a u_a],\\
E'_{\rm cw}  =-\sqrt{R} A_0 [\left( u_0+ \zeta_s u_s \right) + (-1)^f \zeta_a u_a].
\end{eqnarray}
The field leaving the light port is given by $E_L = -\sqrt{R} E'_{\rm ccw}+\sqrt{T} E'_{\rm cw}$, where the negative sign in the first expression is due to the reflection of the counterclockwise field from a hard boundary at the beam splitter. Similarly, the field leaving the dark port is given by $E_D =  \sqrt{T} E'_{\rm ccw}+\sqrt{R} E'_{\rm cw}$, which can be expanded as
\begin{equation}
E_D = A_0 (T\!-\!R)  (u_0+ \zeta_s u_s)+ A_0 \left( (-1)^g T - (-1)^f R \right)\zeta_a u_a(x,y) \label{EDeqn1}.
\end{equation}
Notice that the components of both the trapping and symmetric scattered fields which exit through the dark port suffer destructive interference due to the prefactor $(T\!-\!R)$, and cancel exactly when $T=R$, which corresponds to perfect interferometer visibility. By contrast, constructive interference can be achieved for the antisymmetric scattered field through an appropriate choice of $g$ and $f$. The term in Eq.~(\ref{EDeqn1}) relating to the antisymmetric scattered field can be simplified by defining $m$ to be the difference in the number of reflections experienced by the clockwise and counterclockwise fields after interaction with the particle, such that $m=f-g$, so that
\begin{equation}
E_D = A_0 (T\!-\!R)  (u_0+ \zeta_s u_s)+ (-1)^g A_0 \left( T - (-1)^{m} R \right)\zeta_a u_a(x,y) .
\end{equation}
In the case that $m$ is odd, the antisymmetric part of the scattered field constructively interferes at the dark port, as shown by the $(T\!+\!R)$ prefactor. If the total number of mirrors in the interferometer is odd, $m$ is odd and this condition is met. It is also apparent that the sign of the antisymmetric coefficient will depend on $g$. The only effect this has is to alter the sign of the detected photocurrent $i_x$, with the sensitivity of the measurement left unchanged. Hence, without loss of generality we set $g=2$ as in Fig.~\ref{layout}, to finally find
\begin{eqnarray}
E_{D, \, m~{\rm odd}} = A_0 (T\!-\!R)  (u_0+ \zeta_s u_s)+ A_0 (T\!+\!R)\zeta_a u_a(x,y) \label{EDeqn2}\\
E_{D, \, m~{\rm even}} = A_0 (T\!-\!R)  (u_0+ \zeta_s u_s)+ A_0 (T\!-\!R) \zeta_a u_a(x,y).
\end{eqnarray}
 SNR enhancement requires constructive interference of the antisymmetric term, and hence $m$ odd. Henceforth we only consider this case, with $A_0$ set to be real without loss of generality. The mean photon number flux reaching each position in the detector is given by $\langle n_D(x,y) \rangle =\frac{\epsilon_0 \lambda}{2 h}E_D^*E_D$, where $h$ is Planck's constant and $\epsilon_0$ is the vacuum permittivity. Substituting Eq.~(\ref{EDeqn2}) into this expression we find
\begin{equation}
\langle n_D \rangle = \frac{\epsilon_0 \lambda}{2 h} A_0^2 u_0 (T\!-\!R)[(T\!-\!R) \left( u_0 +(\zeta_s+\zeta_s^* ) u_s\right) +(T\!+\!R)(\zeta_a+\zeta_a^*) u_a] \label{nDeqn},
\end{equation}
where terms of O(2) in $\zeta_a$ and $ \zeta_s$ have been neglected since $\{ \zeta_a, \zeta_s \} \ll 1$. This is detected on a quadrant detector, and subtraction of the resulting photocurrents is performed in the standard manner to infer the position. We assume that the detector size is large compared to the beam size. The sum and difference photocurrents are then given by
\begin{equation}
\langle i_N \rangle =\iint \limits_{-\infty -\infty} ^{ \infty ~ \infty}  \! \! \langle n_D(x,y)\rangle dx \, dy
\end{equation}
and
\begin{equation}
\langle i_x \rangle = \iint \limits_{-\infty ~0} ^{ \infty ~ \infty}  \! \! \langle n_D(x,y)\rangle dx \, dy - \iint \limits_{-\infty -\infty} ^{ \infty ~ 0} \!\! \langle n_D(x,y)\rangle dx \, dy ,
\end{equation}
where the photocurrents $\langle i_N \rangle$ and $\langle i_x \rangle$ are in units of electrons per second. $\langle i_N \rangle$ is the mean total photocurrent generated by the light hitting the detector. This also gives the shot noise variance $\Delta^2 i_N$, which for shot noise limited detection is the noise on the position measurement. Using the normalization property of mode functions given in Eq.~(\ref{mode}), and now neglecting terms of O(1) in $\zeta_a$ and $ \zeta_s$, we find the mean total photocurrent
\begin{equation}
\Delta^2 i_N=\langle i_N \rangle= \frac{\epsilon_0 \lambda}{2 h}(T\!-\!R) ^2 A_0^2. \label{iN}
\end{equation}

\begin{figure}
 \begin{center}
   \includegraphics[width=7cm]{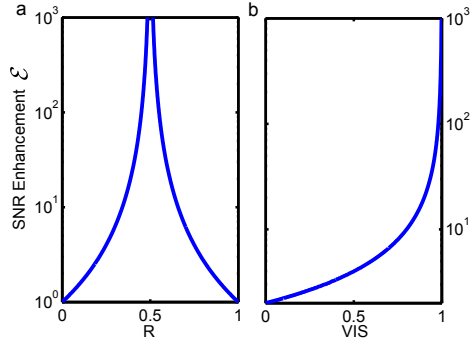}
   \caption{SNR enhancement of the Sagnac interferometer over direct detection, as a function of ({\bf a}) beam-splitter reflectivity and ({\bf b}) Sagnac interferometer visibility. For ({\bf a}), $T+R=1$ has been used, which assumes there is no loss in the Sagnac beam splitter.}
 \label{plots} 
 \end{center}

\end{figure}

The mean photocurrent difference can be found in a similar manner. Since it is obtained by subtracting the flux on one half of the detector from that on the other, the intrinsically symmetric terms $u_0^2$ and $u_0 u_s$ in Eq.~(\ref{nDeqn}) can be ignored. The result is that
\begin{equation}
\langle i_x \rangle = \frac{\epsilon_0 \lambda}{2 h} (T\!-\!R)(T\!+\!R)(\zeta_a+\zeta_a^*) A_0^2 \left( \iint \limits_{-\infty ~0} ^{ \infty ~ \infty}  \! \! u_a u_0 dx \, dy -\iint \limits_{-\infty -\infty} ^{ \infty ~ 0} \!\! u_a u_0 dx \, dy\right) \label{ixEqn}.
\end{equation}
Due to the symmetry of $u_0$ and $u_a$, this simplifies to
\begin{equation}
\langle i_x \rangle = \frac{\epsilon_0 \lambda}{h} (T\!-\!R)(T\!+\!R)(\zeta_a+\zeta_a^*) A_0^2  \iint \limits_{-\infty ~0} ^{ \infty ~ \infty}  \! \! u_a u_0 dx \, dy  \label{ixEqn2}.
\end{equation}
We can define an overlap integral $\eta_{a0}$ as
\begin{equation}
\eta_{a0} =2 \iint \limits_{-\infty ~0} ^{ \infty ~ \infty}  \! \! u_a u_0 dx \, dy ,
\end{equation}
which quantifies the similarity between the mode containing particle position information $u_a$, and the detected mode, given by ${\rm sign}(x) \times u_0(x,y)$ \cite{Tay}. Substituting this into Eq.~(\ref{ixEqn2}) we find
\begin{equation}
\langle i_x \rangle = \frac{\epsilon_0 \lambda}{2 h} (T\!-\!R)(T\!+\!R)(\zeta_a+\zeta_a^*) A_0^2 \eta_{a0}.
\end{equation}
 Using this expression and Eq.~(\ref{iN}) for the shot noise variance, we find the shot noise limited SNR for particle tracking in the $x$ direction to be
\begin{equation}
{\rm SNR}_x = \frac{\langle i_x \rangle ^2}{\Delta^2 i_N} = \frac{\epsilon_0 \lambda}{2 h}(T\!+\!R)^2(\zeta_a+\zeta_a^*)^2 A_0^2 \eta_{a0}^2 \label{SNRx}.
\end{equation}
   An identical result also follows from a quantum treatment of the optical fields, with the shot noise being the result of vacuum noise entering the dark port of the interferometer, rather than being phenomenologically included with the assumption of Poissonian statistics upon detection. It is useful to compare this result to the SNR achieved with standard direct detection. In that scheme, for an input field of $E_0$ we get an output field of $E_0 +E_p$ after the optical tweezers. However, since typical trapping powers are of the order 1 W (for example see Ref.~\cite{Ashkin1987}), and typical photodiodes used for detection have saturation thresholds below 10 mW \cite{Thorlabs}, the optical field is attenuated prior to detection.  To enable a fair comparison, we attenuate the light so that it is the same brightness as the interferometer case, finding
\begin{eqnarray}
E^{{\rm dir}}_D &=& (T\!-\!R) (E_0+ E_p)\\
&=& (T\!-\!R) (E_0+ E_s + E_a),
\end{eqnarray}
so that
\begin{equation}
\Delta^2 i^{{\rm dir}}_N = \frac{\epsilon_0 \lambda}{2 h}(T-R)^2 A_0^2 = \Delta^2 i_N
\end{equation}
and
\begin{equation}
\langle i^{{\rm dir}}_x \rangle = \frac{\epsilon_0 \lambda}{2 h}(T-R)^2 (\zeta_a+\zeta_a^*)^2 A_0^2 \eta_{a0} ,
\end{equation}
where we have again set $A_0$ to be real without loss of generality. This results in a SNR of 
\begin{equation}
{\rm SNR}^{{\rm dir}}_x = \frac{\epsilon_0 \lambda}{2 h}(T\!-\!R)^2(\zeta_a+\zeta_a^*)^2 A_0^2 \eta_{a0}^2
\end{equation}
which is identical in form to the Sagnac SNR except that the $(T\!+\!R)^2$ term in Eq.~(\ref{SNRx}) becomes $(T\!-\!R)^2$ here, substantially degrading the SNR when $T\!\! \approx \!\! R$. Explicitly, the SNR enhancement factor ${\mathcal E}$ for the Sagnac over direct detection is
\begin{equation}
{\mathcal E}=\frac{{\rm SNR}_x}{{\rm SNR}^{dir}_x}=\frac{(T\!+\!R)^2}{(T\!-\!R)^2} \label{Enhance1}
\end{equation}
which is shown as a function of $R$ in Fig.~\ref{plots}~{\bf a}, assuming a loss-less beam splitter such that $T=1-R$. Note that ${\mathcal E}$ tends to infinity as $(T\!-\!R)$ goes to zero. This is clearly unphysical since it corresponds to perfect interference on the Sagnac beam splitter, which requires perfect polarization and spatial overlap as well as $R=T$.  A physically useful parameter which includes all non-ideal effects is the interferometer visibility VIS, given by
\begin{equation}
 {\rm VIS} = \frac{\langle n_L \rangle - \langle n_D \rangle }{\langle n_L \rangle + \langle n_D \rangle } = 1-2 \frac{(T\!-\!R)^2}{(T\!+\!R)^2}  .
\end{equation}
 The visibility quantifies the mode-overlap between the two beams in the interferometer, with a visibility of 1 indicating perfect mode matching. Using this and Eq.~(\ref{Enhance1}), we can express the enhancement factor in terms of the visibility as
\begin{equation}
{\mathcal E}=\frac{2}{1-{\rm VIS}} \label{Evis} .
\end{equation}
The enhancement ${\mathcal E}$ as a function of VIS is shown in Fig.~\ref{plots}~{\bf b}. We see that as the mode overlap goes to unity, the enhancement again approaches infinity.

 In reality the enhancement is limited by the ratio of the trapping field intensity to the detector saturation threshold. In order to compare Sagnac interferometer detection to standard detection, the optical intensity in standard detection is attenuated by a factor of $(T\!-\!R)^2$. However, once the optical power is below the saturation threshold of the detector, it is no longer sensible to apply more attenuation. Once this limit has been reached, no further advantage can be had from improving the visibility of the Sagnac, since there is no requirement to further reduce the optical power reaching the detector.

Practically, the maximum enhancement conferred by the Sagnac interferometer is achieved when it is used to reduce the detected light intensity to the point that the total photocurrent defined in Eq.~(\ref{iN}) is just within the saturation threshold $\langle i_{\rm sat} \rangle$,
\begin{equation}
\langle i_{\rm N} \rangle = \frac{\epsilon_0 \lambda}{2 h}(T\!-\!R) ^2 A_0^2 =   \langle i_{\rm sat} \rangle \label{saturation}.
\end{equation}
Rearranging this, we find
\begin{equation}
(T\!-\!R) ^2 = \frac{2 h}{\epsilon_0 \lambda } \frac{\langle i_{\rm sat} \rangle}{A_0^2}  ,
\end{equation}
which can be substituted into Eq.~(\ref{Enhance1}) to find the maximum enhancement of
\begin{equation}
{\mathcal E}_{\rm max} =  \frac{\epsilon_0 \lambda }{2 h } \frac{A_0^2}{ \langle i_{\rm sat} \rangle} =\frac{\langle i_{N~0}\rangle }{ \langle i_{\rm sat} \rangle}, 
\end{equation}
where $\langle i_{N~0}\rangle$ is the mean photocurrent that would result from the trapping field if there was no attenuation, and we have assumed the Sagnac beam splitter is loss-less, so $T+R=1$.  This limit will depend on the trapping field intensity and the specific detector used. Supposing a Thorlabs PDQ30C quadrant detector was used with a trapping intensity of 1~W as in Ref.~\cite{Ashkin1987}, the maximum enhancement would be approximately 1000.

 To assess the usefulness of this technique we consider a specific example. If optical tweezers are set up in a Sagnac interferometer with visibility of $99\%$, Eq.~(\ref{Evis}) indicates that the SNR would be enhanced by a factor of 200. As shown in Eq.~(\ref{SNRx}), the SNR is proportional to the real part of the antisymmetric scattered field intensity, given by $(\zeta_a+\zeta_a^*)^2 A_0^2$, and is therefore proportional to $r^6$, where $r$ is the particle radius~\cite{Berne,Hulst}. A 200 times increase in SNR  would therefore allow a reduction in the minimum detectable particle size when compared to standard detection of $200^\frac{1}{6}$, or approximately 2.4 times. The sensitivity could be further improved by using spatial homodyne detection instead of the quadrant detector, as quadrant detection has been shown to perform sub-optimally~\cite{Hsu,Tay}.

Finally, we note that the described configuration of the interferometer will only enhance the $x$ position detection because the interfered beams are only flipped in the $x$ direction on reflection against interferometer mirrors. Reflections of both the $x$ and the $y$ directions are required in order to extend this technique to enhanced $x$-$y$ position detection. This can be achieved with a 3-dimensional layout of mirrors.

\section{Conclusion}

By using a Sagnac interferometric detection scheme, the signal-to-noise ratio for particle tracking in optical tweezers is enhanced by a factor which increases as the interferometer visibility approaches $100\%$, up to a maximum enhancement defined by the ratio of the trapping field intensity to the detector saturation threshold.  This improvement comes about because the interferometric scheme results in destructive interference of the trapping field at the dark port without affecting the information carrying part of the scattered field. 
If optical tweezers were set up in a Sagnac interferometer with visibility of $99\%$, the signal-to-noise ratio would be  enhanced by a factor of 200, which would consequently enable tracking of 2.4 times smaller particles than the equivalent standard optical tweezers scheme.

\section{Acknowledgments}

We thank Sean Simpson for contributions towards experimental feasibility studies of the ideas presented in the paper.
This work was supported by the Australian Research Council Discovery Project Contract No. DP0985078.

\section{References}

\bibliographystyle{jphysicsB}

\begin{thebibliography}{10}
\bibitem{Thorlabs}
For example, the commonly used Thorlabs PDQ30C quadrant detector has a 1 mW saturation threshold.
\end{thebibliography}

\end{document}